\documentclass[a4paper,fleqn]{cas-dc}

\usepackage[numbers]{natbib}

\def\tsc#1{\csdef{#1}{\textsc{\lowercase{#1}}\xspace}}
\tsc{WGM}
\tsc{QE}
\tsc{EP}
\tsc{PMS}
\tsc{BEC}
\tsc{DE}
\begin{document}
\let\WriteBookmarks\relax
\def\floatpagepagefraction{1}
\def\textpagefraction{.001}
%\shorttitle{Leveraging social media news}
%\shortauthors{CV Radhakrishnan et~al.}

\title [mode = title]{Non-resonant subwavelength imaging by dielectric microparticles}                      
%\tnotemark[1,2]

%\tnotetext[1]{This document is the results of the research
%   project funded by the National Science Foundation.}

%\tnotetext[2]{The second title footnote which is a longer text matter
 %  to fill through the whole text width and overflow into
  % another line in the footnotes area of the first page.}

\author[1]{Reza Heydarian}
\cormark[1]
\cortext[cor1]{Corresponding author}
\ead{reza.heydarian@aalto.fi}
\address[1]{Department of Electronics and Nano-Engineering, Aalto University, P.O. Box 15500, FI-00076 Aalto, Finland}

\author[1,2]{Constantin Simovski}
\address[2]{Faculty of Physics and Engineering, ITMO University, 199034, Birzhevaya line 16, Saint-Petersburg, Russia}

\begin{abstract}
 Recently a hypothesis explaining the non-resonant mechanism of subwavelength imaging granted by a dielectric microsphere has been suggested. In accordance to the hypothesis, the far-field image of a subwavelength scatterer strongly coupled to a microsphere by near fields is offered by the scatterer polarization normal to the sphere surface. The radiation of a closely located normally oriented dipole is shaped by the microsphere so that the transmitted wave beam has a practically flat phase front. Then this beam turns out to be imaging i.e. keeps the subwavelength information about the dipole location to a long distance from it. However, this mechanism of subwavelength imaging was only supposed in our previous paper. In this paper, we present a theoretical study which confirms this hypothesis and explain the underlying physics of the imaging
which may hold in several scenarios when either a flat or a slightly diverging phase front of the wave beam formed by a microsphere enables the 
subwavelength resolution of two dipole sources in presence of different focusing lenses. We numerically simulate one of these scenarios -- that when the focusing lens is located rather closely to the beam-forming microsphere -- at the distance of the order of $100\lambda$ -- and represents a microsphere itself. In our simulations we replace a 3D microsphere by a 2D "sphere" (microcylinder) so that to use an available electromagnetic solver for a dielectric microparticle of large optical sizes. The physical mechanism of the imaging does not suffer of this replacement.  
\end{abstract}

%\includegraphics{figs/grabs.pdf}

%\item Research highlights item 1
%\item Research highlights item 2
%\item Research highlights item 3


\maketitle

\section{Introduction}

Nine years ago, capacity of a simple dielectric microsphere to offer a  label-free nanoimaging has been experimentally revealed\cite{Hong}. If the object is complex, e.g. represents a pair of closely located dipole scatterers, a deeply subwavelength resolution $\delta\sim \lambda/5-\lambda/10$ was obtained for a range of refractive index and a broad range of the sphere radii. Since the image of an object comprising the subwavelength details is magnified it can be recorded without post-processing or developed by a usual microscope \cite{Yang,Lecler,Kassamakov,Astratov,Zhou,Maslov,Cang,NC}.

Before 2011 the functionality of the magnified subwavelength imaging without fluorescent labels was known only for the so-called hyperlens -- a tapered/curved nanostructure with alternating plasmonic and dielectric constituents forming the so-called hyperbolic meta-material \cite{Hyper1,Hyper2,Hyper3,Hyper4}. Since a simple microsphere is incomparably cheaper, grants much finer resolution and higher magnification {than} any available hyper-lens, a spherical micro-lens has been commonly recognized the most promising device for in-vivo label-free nanoimaging. It attracted a lot of attention of researchers. However, the mechanisms of its operation have not been fully understood up to now.

Attempts to refer the subwavelength imaging functionality of the microsphere to the resonances of the whispering gallery and multiple Mie-resonances of a spherical cavity were done in works \cite{Yang,Lecler,Zhou,Maslov,Cang}. However, these works referred to special cases and could not explain the majority of experimental results. Those theoretical works which pretended to explain the hyperlens functionality for non-resonant microspheres based on the photonic nanojet phenomenon were later claimed erroneous \cite{Astratov1}.

In our recent work \cite{Reza} the hyperlens-like  operation of a dielectric microsphere beyond the resonances was presumably related to the normal polarization of the scattering object. Two mechanisms of the subwavelength imaging of a dipole polarized normally with respect to the sphere were assumed. In both these mechanisms the sphere forms the so-called imaging beam which carries the information about the source exact location to a long distance from the source.  The difference between two mechanisms assumed in \cite{Reza} is in two very different types of the imaging beam.  One type of the imaging beam implied the paraxial focusing of the beam by the objective lens and the coherent imaging of the object, another one -- the tight focusing and the non-coherent imaging. In the first scenario, the imaging beam was ejected from the equatorial plane of the sphere and represented either one or few (2-3) conical beams. This scenario was supposed in \cite{Reza} for a sphere whose radius $R$ is of the order of $(2-5)\lambda$. In this case, a normally polarized dipole excites in a sphere few creeping waves propagating along the sphere surface and forming under it a standing-wave pattern that can be referred {as the interference pattern} of few TM-polarized leaky modes. The leakage of these modes is the ejection of the corresponding creeping waves from the equator of the sphere. Our numerical simulations in \cite{Reza} have shown: these ejected rays form a divergent beam with TM-polarization that has the alternating phase versus the polar angle and the angular distribution of the intensity resembles the Bessel function. It is not a usual Bessel beam, since it is not the Bessel function of the radial coordinate. However, when we analyzed its evolution up to the distances $500-600\lambda$ from the microparticle center we saw that the angular distribution of the intensity kept the same at all distances from the sphere.Also, in \cite{Reza} we have shown that two mutually coherent dipoles with the phase shift proportional to the gap between them, create an interference pattern having two phase centers (virtual sources) separated by a gap which is optically large and proportional to the gap between the real sources. Unfortunately, we could not simulate the focusing of these two imaging beams and this our concept remained not completely proved.   

Note that in \cite{Reza} we have studied not a microsphere, but a microcylinder. We could not find a reliable tool to simulate optically large spheres. However, in the 2D problem the underlying physics keeps the same, because we do not study the resonant phenomena and all effects in which we concentrate are related solely to the curvature of the dielectric interface. Sure, the cylindrical cross section operates similarly to the spherical one.  We {plan} to finalize the study of the subwavelength imaging by such modestly large spheres in our next works where we will explicitly show the imaging.  

In the present work we concentrate on the second scenario of the imaging supposed in \cite{Reza}. It is based on the creation of a non-divergent or very slightly divergent imaging beam that is possible if the microsphere as more substantial i.e. $R>10\lambda$. For so large spheres the Mie resonances vanish, the leaky modes form a quasi-continuum, and the formation of the imaging beam can be governed by the laws of geometrical optics (GO). Namely, GO can be applied to describe the refraction of the wave on the back side of the sphere. As to the penetration of the dipole radiation into the sphere, it is, on the contrary, a quasi-static effect. A dipole source located on the dielectric interface creates a strong local polarization in its vicinity. A subwavelength domain of maximal dielectric polarization can be united with a dipole into an effective source. It is also a dipole effectively located inside the sphere. Therefore we do not need to consider the diffraction of the dipole radiation by the front side of the sphere. Of course, the diffraction effects would be important if we considered the scattered fields. They would be important also for the transmitted field if we located the primary source at a substantial distance from the sphere. However, for the dipole source located just on the front surface of a large dielectric sphere, the transmitted wave beam is formed by the back half of the spherical surface and it is possible within the frameworks of GO. The paths of all rays to the back side of the sphere are optically large and the surface curvature is small, so, the prerequisites of the GO approximation are respected. 
The problem arises only at the frequencies of whispering-gallery resonances. However, in the range $R>10\lambda$ these resonances are sparse over the frequency axis and we do not consider them. These speculations allowed us to claim that the collimated transmitted beam is possible in the transmission region \cite{Reza}.

\section{Governing idea and its discussion}

\begin{figure}[htbp]
\centering
\includegraphics[width=0.32\textwidth,height=0.29\textwidth]{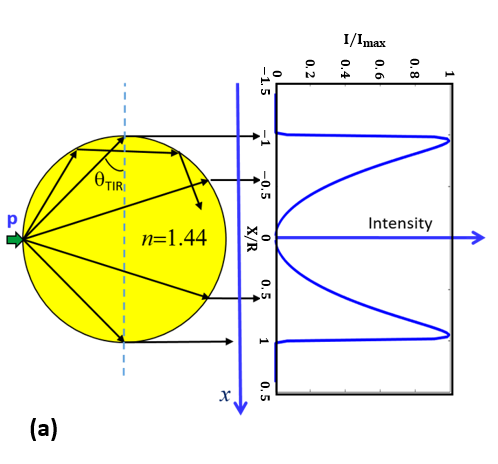}
\includegraphics[width=0.37\textwidth,height=0.26\textwidth]{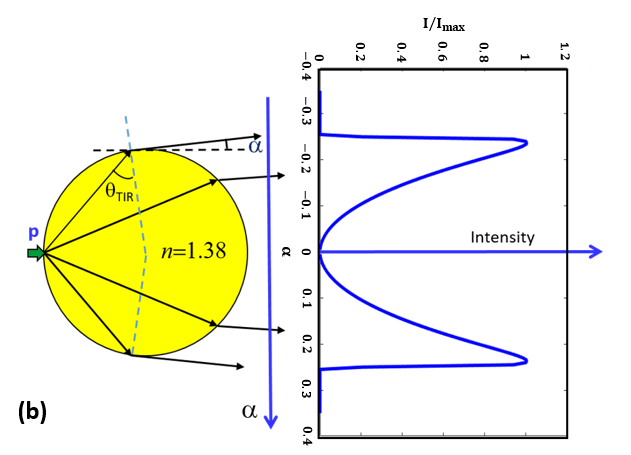}
\caption{A normally polarized dipole produces behind the sphere a transmitted beam depicted in the approximation of GO 
for two values of the refractive index. The rays propagating from the dipole source to the back half of the sphere have $\theta<\theta_{TIR}$ and do not experience total internal refraction. In the case (a) $n=1.44$, and they all refract in parallel to the beam axis $y$ forming a tubular beam.  The phase front of this beam is nearly flat. In the case (b) $n=1.38$, these rays are slightly diverging, and the beam represents a spherical wave with annular pattern. The phase center of this wave is located on the optical axis far behind the real source.}
\label{PicNew0}
\end{figure}

For the convenience of readers let us repeat the speculations of \cite{Reza}.
In the approximation of GO the rays of an internal dipole source refract on the back interface as if it were flat. 
Using the Snell's law it is easy to calculate the direction of rays in the transmitted beam. For a specific value of the refractive index ($n=1.44$) all these rays are parallel to the axis $y$ and the critical angle of total internal reflection, as depicted in Fig.~\ref{PicNew0}(a), is $\theta_{TIR}\approx\pi/4$. In the GO approximation, it is easy to calculate the intensity distribution across the beam. For a normal (radial to the sphere centre) dipole and a tangential one the intensity distributions will be different due to the dipole angular pattern. The result for the intensity distribution in the case of a radial dipole is presented in Fig.~\ref{PicNew0}(a). In accordance to this result, the transmitted beam is tubular. Its polarization is radial with respect to the beam axis -- axis $y$ due to the axial symmetry. In the case when $n<1.44$ the transmitted beam in the GO approximation is not exactly parallel but is slightly divergent. For this case the intensity distribution is shown in Fig.~\ref{PicNew0}(b) as a function of the ray tilt angle $\alpha$. The distribution of $I/I_{max}$ versus $\alpha$ resembles the intensity distribution for a tubular beam versus the radial coordinate $\rho$. 

It is also easy to calculate the phase distribution in these beams. Let the coordinate origin 
$x=y=z=0$ be the center of the sphere. For an internal ray which impinges the back interface of 
the sphere under the incidence angle $\theta$ the optical path from the source point $x=z=0,\, y=-R$ to the plane $y=R$  
equals $\Pi=R[2n\cos\theta +R(1-\cos2\theta)]$. The transverse coordinate of the transmitted ray $\rho=\sqrt{x^2+z^2}$ is equal $R\sin2\theta$, and it is clear that for the case $n=1.44$ (when in accordance to the Snell's law all rays transmit with the zero tilt angle $\alpha=0$ to the axis $y$) the value $k\Pi$ is not exactly constant versus $\rho<R$. However, the phase front of the transmitted beam can be with a suitable accuracy adopted flat because the beam is tubular. The intensity of the beam is effectively concentrated in a ring $R/2<\rho<R$, and inside this ring the phases of all rays in the plane $y=R$ are nearly equal to one another. For example, for $R=10\lambda$ the phase difference between the rays having $\rho=R$ and $\rho=R/2$ is close to $\pi/20$. This difference, though nonzero, is, definitely, more tiny than the error of the GO model. In other words, the GO model of the transmitted beam sketched in Fig.~\ref{PicNew0}(a) is not controversial. When all rays are parallel to one another, the phase font across the beam is almost flat. Similarly, when $n=1.81$ the phase front turns out to be exactly flat in the GO model, whereas the rays inside the ring $R/2<\rho<R$ at the plane $y=R$ will be almost parallel to the beam axis.  
This self-consistency of the model allowed us to hope that such a parallel transmitted beam is really possible. 

Next, for the case $n=1.38$ the GO model delivers the phase distribution which corresponds to the spherical wave and similarly matches the intensity distribution depicted in Fig.~\ref{PicNew0}(b). The surface to which all rays are orthogonal in accordance to the Snell's law is the surface on which all rays of this hollow conical beam have nearly the same phase in accordance to their optical path. In other words, this conical beam is a spherical wave having a point-wise phase center. It is clear that the phase center of this wave is located on the beam axis $y$ very far behind the source. Now we may formulate the governing idea of this work as follows: a large microsphere excited by a point dipole (located on its front surface and polarized normally) forms two types of the wave beam in the region of the transmitted field. The first type of the beam is a hollow non-divergent beam (tubular one), the second type is a spherical wave with annular pattern and very distant phase center. We believe that both these transmitted beams should offer a magnified subwavelength image of a tiny object located on the sphere. We have called both these beams as imaging beams.

\begin{figure}[htbp]
\centering
\includegraphics[width=0.49\textwidth,height=0.28\textwidth]{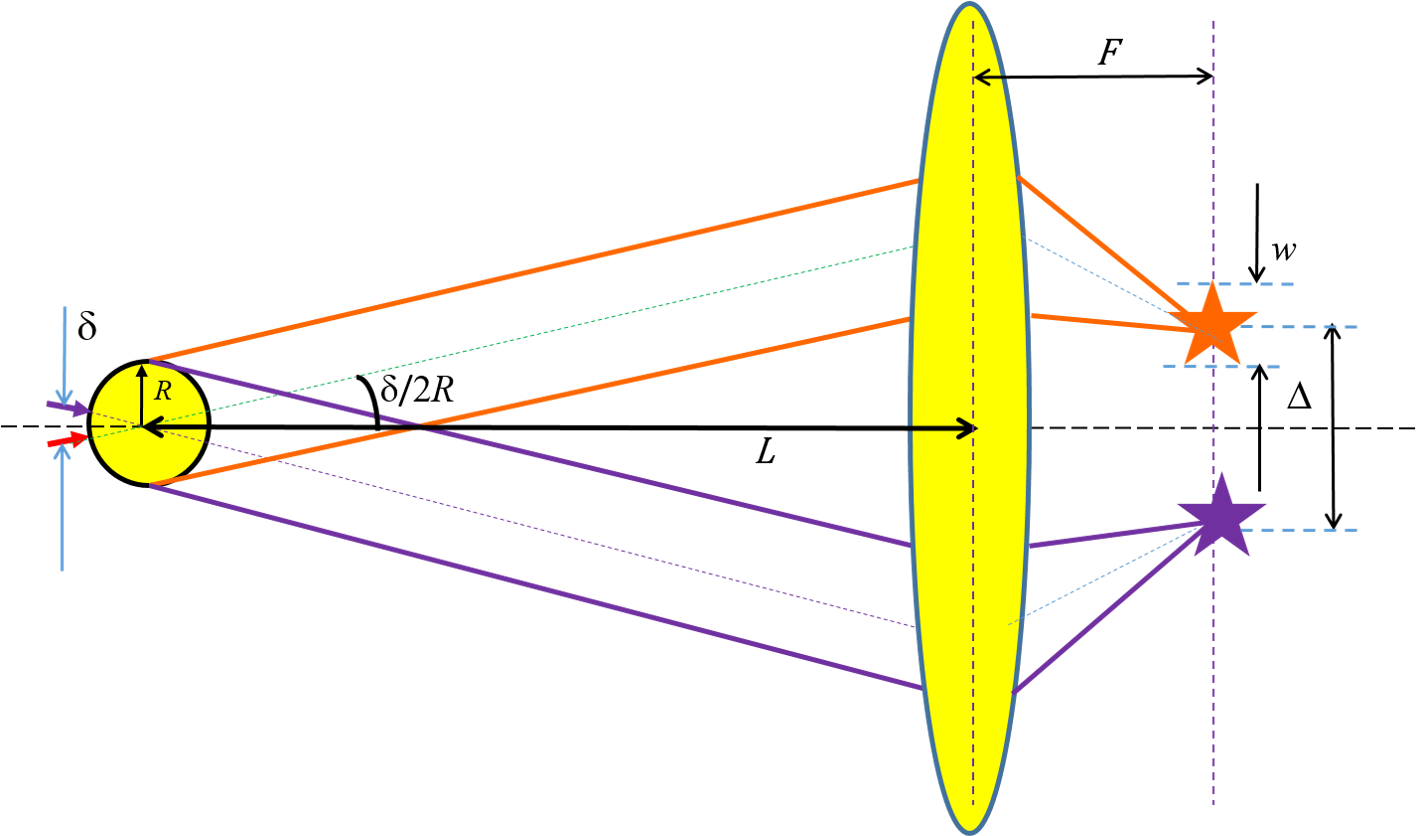}
\caption{Two point dipoles (shown for better visibility by different colors) are located with a subwavelength lateral gap $\delta$ on the front surface of the sphere. The radiations of dipoles are not mutually coherent. They separately produce two non-divergent wave beams whose cross sections are equal to $2R$. These beams create a magnified image of the dual object in the focal plane of the objective lens. Since $\Delta\gg \delta$ the subwavelength size $w<\lambda$ of the focal spot is not needed for superresolution.}
\label{PicNew}
\end{figure}

Let us consider the first scenario of imaging, when a tubular beam with flat phase front propagates along the axis $y$ to the distance much larger than the sphere diameter $2R$. The propagation of such beams in free space and their focusing by lenses are described in works \cite{tour,tour3,radial}. In \cite{tour3} it is pointed out that such beams should offer the spatial superresolution because being focused by a lens with short focal distance (this distance $F$ should be then the value of the order of the beam effective width) this beam converges into a spot of subwavelength size. For this subwavelength focusing 
the beam should be nearly parallel i.e. we need to place the lens at the lower distance from the beam waist than the 
Rayleigh distance $D_R$, also called the diffraction length. It is commonly known that $D_R=\pi R^2/\lambda$, where $R$ is the effective radius of the beam waist (in our case $R$ is the sphere radius). The requirement of the tight focusing is essential because a long-focus lens will collect {our tubular} beam into a ring \cite{radial}. Only a tight focusing grants the strong tilt of the electric field vector in the converging rays with respect to the beam axis and results spot-wise image. In this case the maximum of the intensity is achieved on the beam axis where $\bf E$ is polarized longitudinally.

Though for the first scenario of imaging the tight focusing is required, in this scenario we do not need to obtain the subwavelength focal spot. Two non-divergent beams corresponding to two point dipoles with a small lateral gap $\delta$ between them will create two spots in the focal plane of the objective lens as it is shown in Fig.~\ref{PicNew}. The gap $\delta$ can be deeply subwavelength, the focal spot diameter $w$ can, on the contrary, exceed $\lambda$, but still we will obtain the resolution of two point sources because the distance between the images of our dipoles is drastically enlarged compared to $\delta$. In accordance to GO $\Delta\gg \delta$ if $L\gg 2R$. The maximal spot size $w=\Delta\approx L\delta/R$ corresponds to the ultimate case of the spatial resolution when it is possible that $\delta\ll\lambda\ll w$. All we need for the far-field subwavelength imaging is the negligibly small beam divergence (that holds for $L<D_R$) and the tight focusing so that to image a point by a small spot and not by a ring \cite{radial}.   

As to the imaging in the second scenario, when the transmitted beam is slightly divergent, the superresolution property is evident even without an illustration. Really, let a slightly divergent imaging beam have a point-wise phase center on the axis $y$ where the continuations of the rays shown in Fig.~\ref{PicNew0}(b) cross out. Then a normally polarized dipole located on the front surface of the sphere produces a virtual source located at the left from the sphere at a big distance $Y$ from it ($Y\gg R$). Two dipoles with a tiny gap $\delta$ between them will produce two virtual sources on their axes. The distance $\delta_v$ between them will be  proportional to $\delta$:  $\delta_v=Y\delta/R$. There is no problem to resolve these virtual sources with a microscope if $\delta_v>\lambda$. 

Of course, all these speculations are valid if the GO model is adequate for two types of the imaging beam, but the absence of the internal contradictions in our GO models was inspiring. We made a lot of full-wave numerical simulations for different values of the refractive index $n$ and different size parameters $kR$ of the sphere. In these simulations we found the ranges of $n$ and the ranges of $kR$ which offer both claimed regimes -- that of the tubular beam and that of the hollow conical beam with a far phase center. These simulations are discussed below.

\section{Preliminary simulations}

We performed extensive simulations of 2D "spheres" using COMSOL Multiphysics, which is a software much more rapid and reliable than the other electromagnetic solvers for the 2D problems. 
We have calculated the intensity plots and instantaneous wave pictures of the transmitted beam varying $R$ in the interval $(10-40)\lambda$, where we put $\lambda=550$ nm. We varied $n$ in the interval $1.4-1.7$ typical for optical glasses. We also varied the gap between the "dipole" and the "sphere" from zero to {$\lambda/10$} and found that this gap does not influence the field distributions and only may change the magnitudes. 

Though the simplistic model represented by the plot in Fig.~\ref{PicNew0}(a) does not give a quantitative agreement with our full-wave simulations, it granted a qualitative insight and useful estimates. 
The parallel transmitted beam was obtained for $n=1.49-1.7$ when $R=15-25\lambda$. 
For each $R$ from this range there is band of $n$ for which the transmitted beam is parallel, e.g. for . 
$R=20\lambda$ the parallel beam is formed if $n=1.55-1.62$. When $R=25-30\lambda$ the beam for the same values of $n$ is slightly diverging and corresponds to the type of the imaging beam depicted in Fig.~\ref{PicNew0}(b). For $n=1.40-1.48$ the parallel transmitted beam was not obtained for any radius from the whole range $R=10-40\lambda$. In this case, when $R>20\lambda$ the beam is slightly diverging as in Fig.~\ref{PicNew0}(b). When $10\lambda<R<20\lambda$ and $n=1.40-1.48$ the phase center of the diverging beam is not point-wise and the second scenario of the imaging is not possible.

Further, we have checked how the imaging beams of both types vary at the long distances from the sphere. The replacement of a 3D sphere by a 2D "sphere" (cylinder), when then the point dipole is replaced by a dipole line, allowed us to reliably simulate not only very large microparticles ($R=10-40\lambda$), but also to follow the transmitted beams at very large distances -- hundreds of $\lambda$. Moreover, this replacement allowed us also to simulate the practical case when the dipole source is sandwiched between the sphere and the substrate. Let us stress that this replacement is justified: we study the non-resonant regimes, when the substitution of the 3D sphere by a 2D "sphere" keeps valid the underlying physics. Even numerically there is the good correlation with the 3D case. For example in our 2D simulations the beam effective width  
calculated for $R=20\lambda$ increases by nearly $10$\% at the distance $y\approx 200\lambda$.
The same result corresponds to the evolution of the axially symmetric radially polarized beam  
of the same waist radius $R$ in accordance to formulas of \cite{radial}.  

Unfortunately, we could not implement in our simulations neither the first scenario of the imaging illustrated by Fig.~\ref{PicNew} nor the second scenario. The first scenario cannot be implemented with a realistic microscope. 
The spheres with the radius $R>30-40\lambda$ do not offer the subwavelength imaging
in the experiments and do not form imaging beams in our 2D simulations. However, a microscope objective lens cannot 
have the focus as short as $F<30-40\lambda$. As to the second imaging scenario, its implementation would require to simulate the beam evolution on the paths exceeding $1000\lambda$, that is not possible even in the 2D variant for any available electromagnetic solver.

\begin{figure}[htbp]
\centering
\includegraphics[width=0.52\textwidth,height=0.32\textwidth]{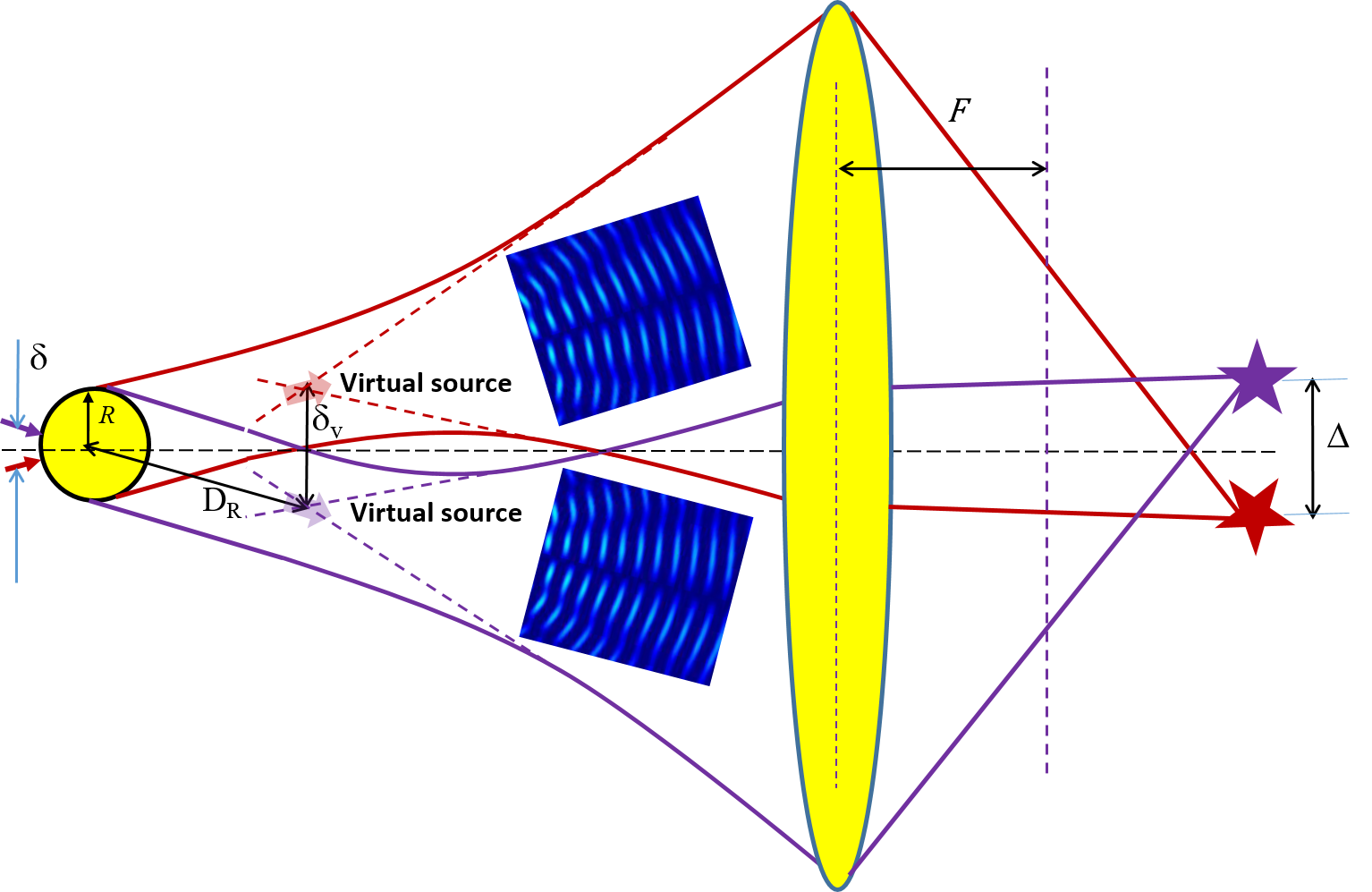}
\caption{Two point dipoles (shown for better visibility by different colors) are located with a subwavelength lateral gap $\delta$ on the front surface of the sphere. The dipoles produce two wave beams whose cross sections are initially equal to $2R$
but at the distance $D_R$ become nearly spherical waves. Though the phase centers of these waves are spread to the area 
of the size $2R$, these areas do not intersect because $\delta_v\gg 2R$. These phase centers can be guessed in the insets (COMSOL simulations) and treated as virtual images of the real sources because $\delta_v$ is proportional to $\delta$. 
}
\label{Picture12}
\end{figure}

However, the analysis of the parallel beam evolution at large distances $y\sim D_R$ (Rayleigh range)
allowed us to reveal the third scenario of subwavelength imaging granted by the microsphere. At these distances 
the intensity starts to oscillate across the beam (i.e. the tubular beam splits onto several hollow beams)   
and simultaneously the tubular beams start to diverge becoming conical ones. At the distance $y\approx D_R$ the beam produced by the sphere $R=20\lambda$ with $n=1.6$ (which looks non-divergent up to 200$\lambda$) transforms into a nearly spherical wave
with the annular pattern. It is not an ideal spherical wave -- its phase center is not point-wise. However, it is an area whose size is smaller than $R$. Therefore, even this spread phase center grants the superresolution. This scenario of imaging is illustrated by Fig.~\ref{Picture12}. The size of the spread phase center can be guessed on the insets (the wave picture of the initially parallel transmitted beam at the Rayleigh range of distances). It can be easily estimated that the wave really diverges from the area whose longitudinal and transverse sizes are smaller than the initial beam radius $R$. These areas can be then treated as two virtual sources formed in the places where two initially parallel beams created by two dipole sources transform into two spherical waves. These virtual sources do not overlap if their lateral size is smaller than the distance $\delta_v$ between them, where $\delta_v=\delta D_R/R=\pi\delta R/\lambda$. The resolution will be finer than $\lambda/\pi$ because the 
lateral size of a virtual source is smaller than $R$ and because a partial overlapping of the virtual sources still allows the microscope to resolve them. 

\begin{figure}[htbp]
\centering
\includegraphics[width=0.47\textwidth,height=0.28\textwidth]{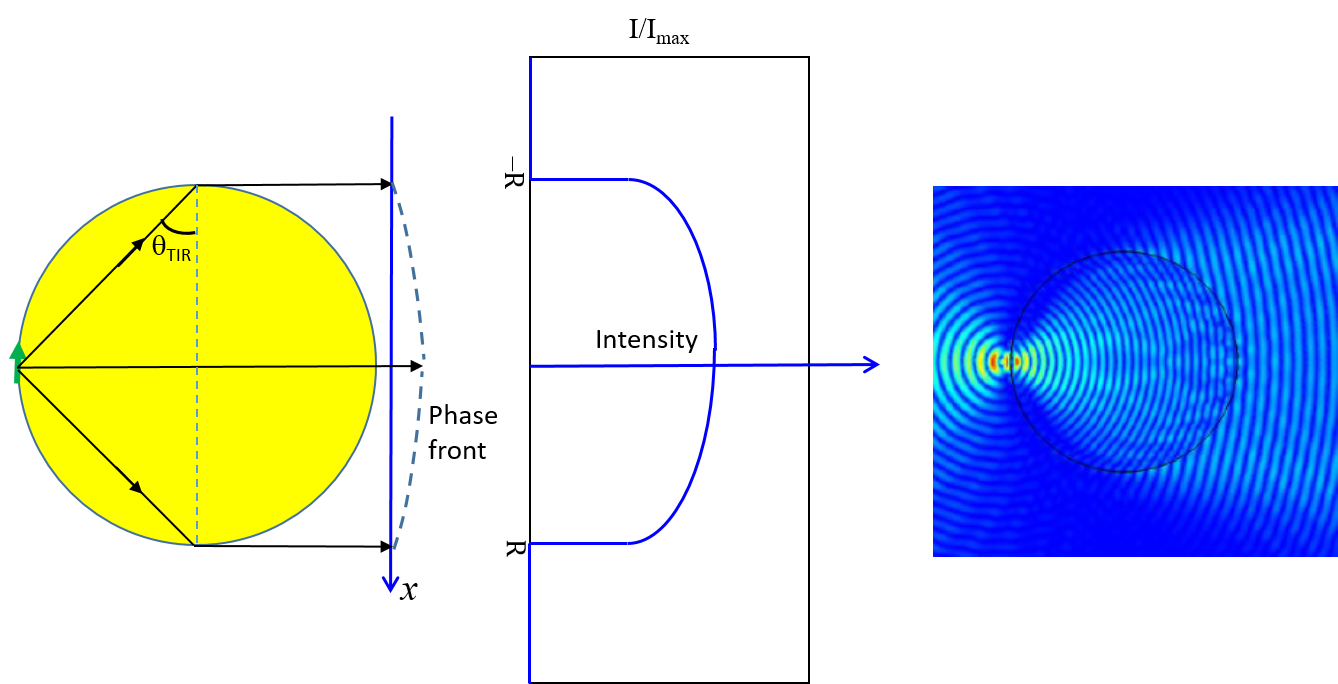}
\caption{A tangentially polarized dipole produces behind the sphere a transmitted beam depicted in the approximation of GO 
for $n=1.44$. The parallelism of the rays contradicts to the curvature of the phase front. In the inset the instantaneous wave picture of the beam with the smallest obtained front curvature is shown. It corresponds to $n=1.7$, $R=5\lambda$}. 
\label{Picture11}
\end{figure}

Now, let us discuss the role of the normal polarization of the dipole in the imaging. If a normal dipole can create the imaging beam why a tangential dipole cannot? The answer becomes clear inspecting Fig.~\ref{Picture11}. The pattern of the tangential dipole results in the intensity distribution having the maximum on the beam axis. For a hollow beam we 
could neglect the curvature of the phase front, for a cylindrical beam it is not allowed. From one side, for $n=1.44$ the rays of the transmitted beam are orthogonal to the plane $(x-z)$, from other side, the phase front has the non-negligible curvature over the beam cross section. For $R=20\lambda$ the phase of the rays with the radial coordinate $\rho=R$ differs from the phase of the central ray in the plane $y=R$ by nearly $\pi/4$ that is not a negligible phase shift. If we change $n$ so that the GO model delivers the exactly flat phase front ($n\approx 1.81$), the same GO model gives the transmitted beam of intersecting rays noticeably tilted with respect to the axis $y$. So, in this case the GO model turns out to be not self-consistent. Therefore we had a little hope to form a parallel beam with a tangential dipole. In our simulations any combinations of $n$ and $kR$ resulted neither in a parallel beam like Fig.~\ref{PicNew0}(a) nor a diverging beam like Fig.~\ref{PicNew0}(b).   
In the inset of Fig.~\ref{Picture11} we show the typical wave picture of the transmitted wave beam for the tangential dipole. 
The curvature of the phase front is varying across the beam and does not correspond to a spherical wave with point-wise or at least subwavelength small phase center. This our result fits that of \cite{Astratov1} where the simulations were performed using the exact solution of the 2D problem (dipole line field diffraction by a cylinder). Though we have a little hope to achieve the subwavelength imaging of a tangentially polarized object, these speculations are, of course, not strict and the principal possibility to obtain the imaging beam cannot be proved involving the GO model and simulations of some particular cases. 
He we may claim only that in our numerical studies we have not found the regime of the imaging beam for a tangential dipole and this negative result fits our theoretical expectations.

\section{Fourth scenario of the subwavelength imaging by a microsphere}

We believe: what an experimenter sees in the microscope looking at two point-wise scatterers through a glass microsphere is the image of the virtual sources created by the imaging beams as it is shown in Fig.~\ref{Picture12}. However, the doubts are possible even after these our simulations because we have not simulated the whole system depicted in this Figure i.e. have not proved our claim completely. A fully convincing explanation of the non-resonant hyperlens-like action of a glass microsphere is a challenging task. In order to publish a scientific paper treating this phenomenon we should have shown namely the superresolution, at least in the numerical simulations. What we have shown were only prerequisites of the superresolution. 

Since, we are unable to simulate the beam evolution in the scale of thousands of wavelengths as it implied in Fig.~\ref{Picture12}, we decided to study the fourth scenario of the superresolution granted by a microsphere. 
This is a modification of our first scenario with the replacement of the macroscopic focusing lens by a microlens. 
The center of the microlens is located at a modest distance $y<100\lambda$ from the object and the microlens has also a modest optical size. This restriction of the simulation area makes the simulation rapid and allows us to perform extensive simulations and optimization of the structure. Unfortunately at the distances $y<100\lambda$ two imaging beams with the waist widths $2R=20-60\lambda$ created by two point dipoles separated by a small gap $\delta<0.5\lambda$ still intersect. Therefore, we cannot obtain a strong magnification of the image -- our $\Delta$ turns our to be of the order of $2\delta$. However, in order to obtain the subwavelength resolution we may utilize the property of a radially polarized non-divergent beam -- the possibility to be focused into a subwavelength spot \cite{tour3,radial}. In our conceptual proof, the microlens is a glass sphere, similar to that which forms the imaging beam. We have seen in our simulations that it may really focus the axially symmetric beam (a parallel one and even a slightly diverging one) into a subwavelength spot. 
  
In order to focus the parallel or slightly diverging beam we did not need a large microlens. It is the "sphere" of the same radius whose center is located at the distances $50-100\lambda$ from the center of the first  "sphere". On the first stage we simulate the subwavelength image of a single dipole source located on the front size of the first "sphere". On the next stage, we replace one dipole source by two non-interfering dipoles with a subwavelength gap between them. Finally, we add a large dielectric block emulating the substrate so that the imaged object is sandwiched between it and the sphere. Repeating all the steps we study the impact of the substrate to the imaging and find that this impact is positive: the substrate plays an important role in the proper formation of the imaging beam.

\section{Subwavelength imaging of a single point dipole}

\begin{figure}[htbp]
\centering

\includegraphics[width=0.51\textwidth,height=0.21\textwidth]{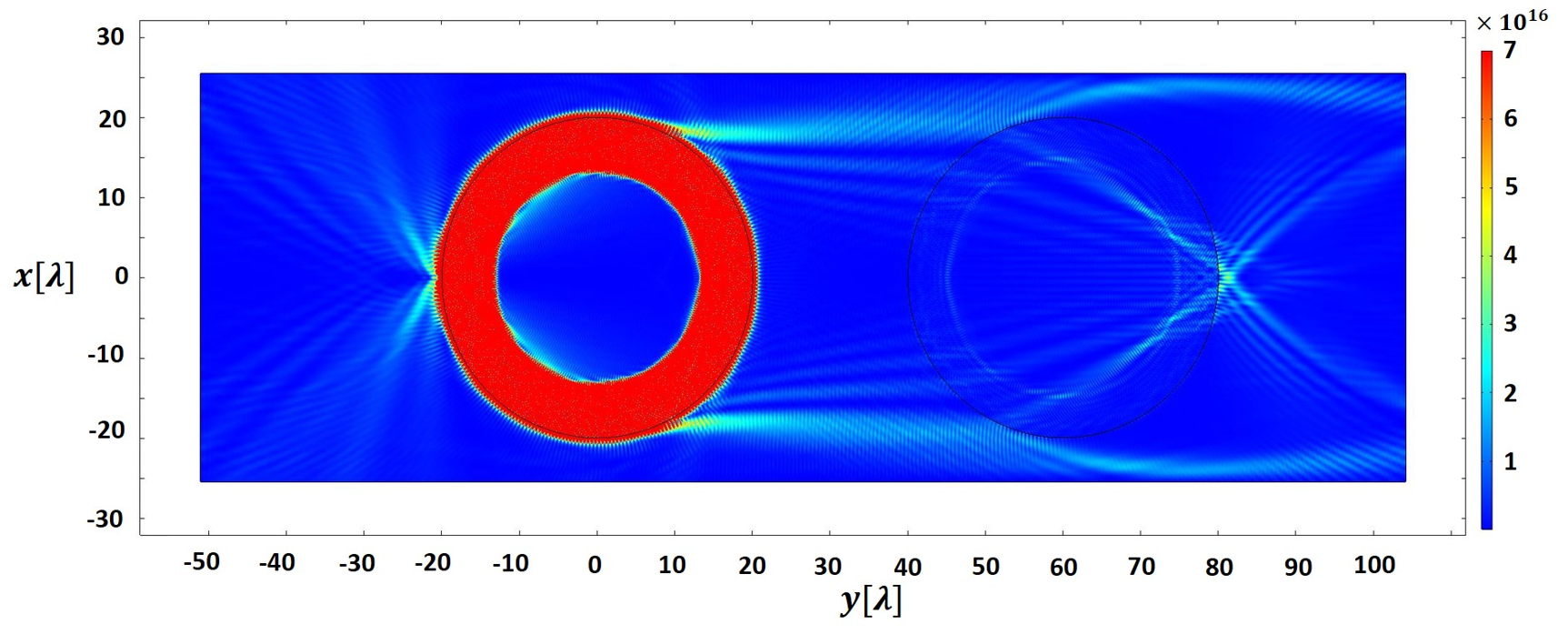}
\caption{Color map {of electric intensity} in the imaging system $R_1=R_2=20\lambda$ excited by a dipole located at a point 
{$x=0,\, y=-R-\lambda/10$}. 
}
\label{Pic1}
\end{figure}

All calculations reported below {refer to} the gap {$\lambda/10$} between the normally oriented Hertzian dipole and the first "sphere". On the stage reported in this section we simulate the image of one dipole source. Recall that in our current scenario we need the subwavelength size of the focal spot $d$.         
Note, that the subwavelength focus in work \cite{tour3} was obtained for the 3D case. 
In \cite{JOSAB} we have theoretically obtained the similar result for the 2D case. 
After extensive simulations we have seen that the optimal radius of the second "sphere" which would grant us the minimal possible $d$ for the case of the parallel imaging beam is equal $R$. Thus our imaging system is a pair of microcylinders of the same radius $R_1=R_2=R$. In Fig.~\ref{Pic1} we see an example of {such imaging} system and the color map of {electric field intensity}. The imaging system consists of two glass "spheres" of radius $R=20\lambda$. The refractive index of the first "sphere" is {$n_1=1.6$} and that of the second "sphere" is $n_2=1.52$. 

This color map represents the intensity distribution that in general corresponds to Fig.~\ref{PicNew0}(a). The differences are 
weak intensity maxima located between two "spheres" in the area $|x|<R$ and weak interference pattern in the imaging beam. 
Both these features result from the multiple reflection of the imaging beam
between two "spheres". If the distance between the "spheres" {increases from $(50-100)\lambda$ to $(150-200)\lambda$,} these features vanish. However, the multiple reflection of the imaging beam between two "spheres" whose centers are distanced by $60\lambda$, is a weak effect, and we saw that it is not harmful for the imaging. 
Therefore, in this paper we report namely this compact structure.      

In the color map we see the image of a point dipole. It is a small bright spot slightly but noticeably distant from the rear edge of the second "sphere". The $y$-coordinate of this {spot center determines the focal plane} of the second "sphere" operating {as a microlens.} Analysing the intensity distribution $I(x)$ in this plane we find the effective width $w$ of this focal spot. {According to the Rayleigh criterion ($0.7$ of the intensity at the local maximum), effective width for the present case is $w=0.37\lambda$.}

\begin{figure}[htbp]
\centering
\includegraphics[width=0.37\textwidth,height=0.29\textwidth]{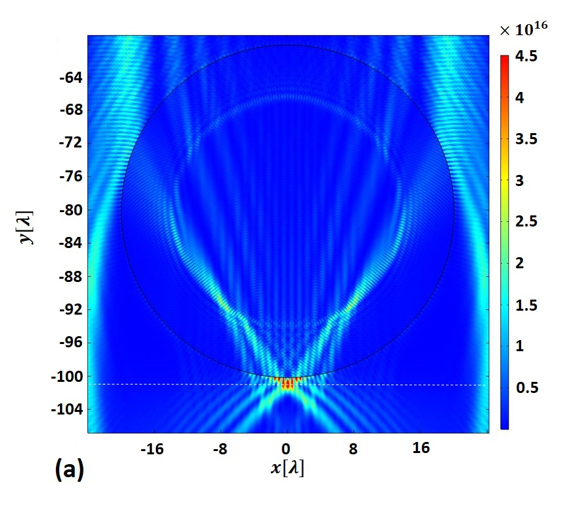}
\includegraphics[width=0.37\textwidth,height=0.29\textwidth]{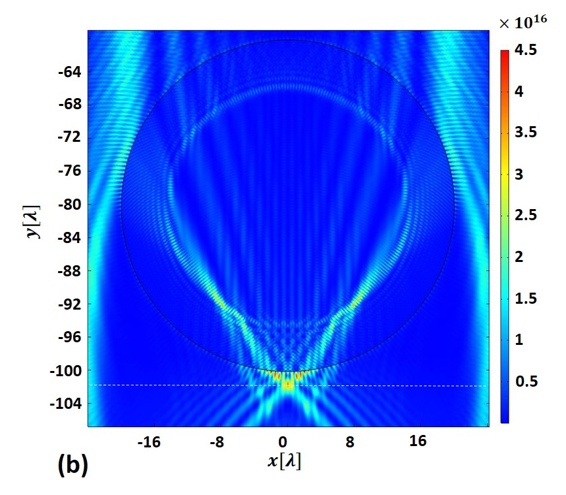}
\caption{ Color map of the intensity in the area of the second "sphere" (microlens) for $n_2=1.55$ (a) and $n_2=1.49$ (b). 
White dashed line marks the microlens focal plane. 
}
\label{Pic2}
\end{figure}

The ratio $w/\lambda$ can be reduced for given $R$ and $n_1$ if we vary the refractive index $n_2$ of the microlens. To understand the impact of $n_2$
we inspected {zoomed} color {maps} of intensity {in the area} of the second "sphere". In Fig.~\ref{Pic2} we {depict these maps for two values of $n_2$.} In both cases, there are noticeable side-lobes. However, for {$n_2=1.55$} the central maximum is more pronounced 
than in the case {$n_2=1.49$}. It shows that for given $R$ and $n_1$ there is an optimal interval of $n_2$. How broad is this band?  

\begin{figure}[htbp]
\centering
\includegraphics[width=0.45\textwidth,height=0.29\textwidth]{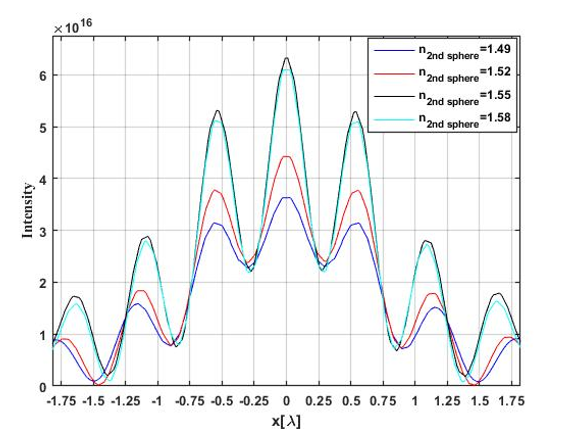}
\caption{Intensity plots in the focal plane of the second "sphere" for four values of the refractive index $n_2$.}
\label{Pic3}
\end{figure}

In {Fig.~\ref{Pic3} we have depicted} the intensity plots calculated in the focal plane of the second "sphere" for several values of $n_2$ varying from $1.49$ to $1.58$ with the step $0.3$. The effective width of the central spot varies from $0.22\lambda$ for $n_2=1.58$ to $0.41\lambda$ for $n_2=1.49$. For smaller $n_2$ the subwavelength imaging of the point source is lost -- the central spot exceeds $\lambda/2$. For $n_2>1.58$ the ratio $w/\lambda$ keeps close to $\lambda/5$, i.e. is deeply subwavelength. For $R=20\lambda$ and $n_1=1.6$ the optimal values of $n_2$ lie within the interval    
$n_2=1.55-1.7$. 

To illustrate  the impact of the optical size of the second (imaging) "sphere", we show our best result obtained for the case when the imaging beam is formed by a "sphere" of radius $R=30\lambda$. In this case our imaging system also comprises two "spheres" of the same radius. The optimal refractive index of the imaging "sphere" is $n_1=1.6-1.7$ and the same refers to the optimal values of $n_2$. The imaging beam produced by the first "sphere" in the case $n_1=n_2=1.6$ is slightly diverging and $I(\alpha)$, calculated in the absence of the second "sphere", qualitatively mimics that shown in Fig.~\ref{PicNew0}(b). 

\begin{figure}[htbp]
\centering

\includegraphics[width=0.5\textwidth,height=0.25\textwidth]{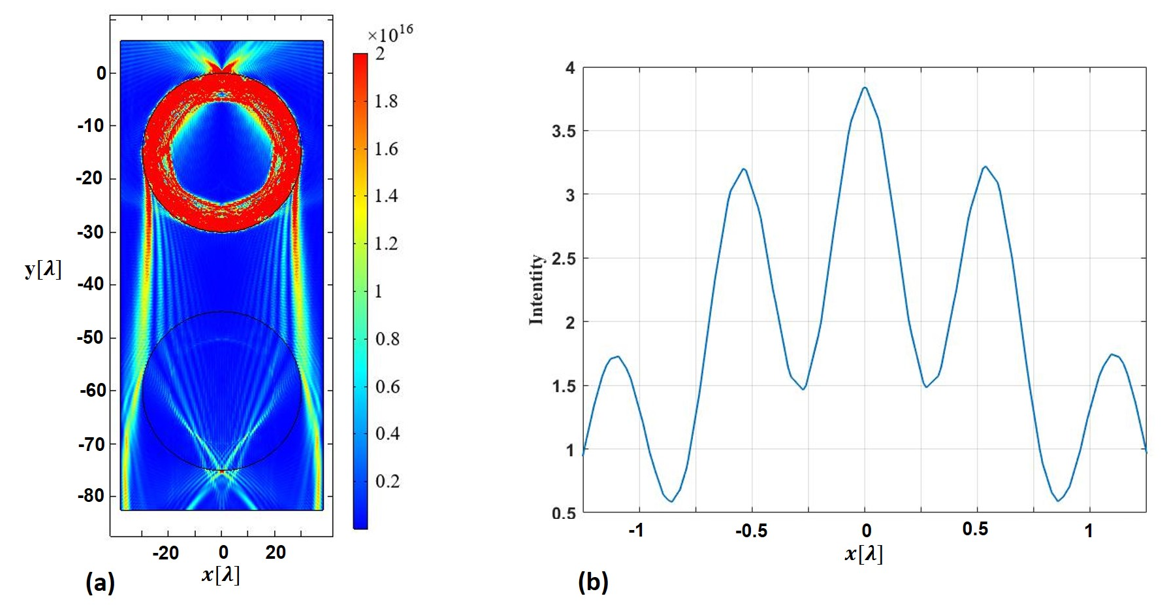}
\caption{Color map of the electric intensity in the imaging system $R_1=R_2=30\lambda$, $n_1=n_2=1.6$ (a) and intensity plot  
in the focal plane (b).
}
\label{PicNew1}
\end{figure}

If the second "sphere" is added at the center-to-center distance {$90\lambda$} from the first one, the features of the parasitic interference are more pronounced than {those we have observed} in the case of the parallel {beams} for the same distance between the spheres. Increasing this distance we have checked that this parasitic effect has no impact to our imaging. 
Therefore, we may center the second sphere at the modest distance {$90\lambda$} and show the whole imaging system 
in the same color map as we did for the parallel beam. This color map for the slightly diverging beam is presented in Fig.~\ref{PicNew1}. 
In this map we see the similar image as we observed in the case of the parallel beam. 
In accordance to the intensity plot, calculated in the focal plane and depicted in Fig.~\ref{PicNew1}(b), in the present case 
we have $w=0.308\lambda$. This result cannot be noticeably improved under restriction $n_1=n_2<1.7$.

For a glass "sphere" there is an interval of $R/\lambda$ for which the achievable ratio $w/\lambda$ is deeply subwavelength. 
In our study, when the dielectric material is an optical glass the optimal values of the ratio $R/\lambda$ lie around $20$. Our study explains why the non-resonant hyperlens functionality cannot exist for very large dielectric spheres as well for submicron spheres. There is the range of size parameters for which the phenomenon exists. It is rather broad but not ultra-broad.
This our result fits the experimental observations of \cite{Hong}: in the incoherent illumination the best lateral resolution of two point scatterers ($\lambda/6-\lambda/8$) was achieved for a glass sphere ($n=1.46$) whose the radius {was} within the interval $2.5\lambda\le R\le 12\lambda$.
In our 2D case {the corresponding interval is} $10\lambda\le R\le 30\lambda$.

From these studies we may conclude that both regimes of the parallel and slightly diverging imaging beams are suitable for the 
fourth scenario of subwavelength imaging -- that by two microsphere. How fine is this image, depends on the optical size of the imaging "sphere" and on its refractive index.  

Returning to our simulations, we should point out that the interval $R/\lambda=(15-25)$ is optimal for sobwavelength imaging of a point source. In this interval the condition 
$n_1=1.49-1.7$ allows us to observe the parallel beam similar to that depicted in Fig.~\ref{PicNew0}(a). For smaller $n_1$ 
($n< 1.49$) the imaging beam becomes slightly divergent. In this case the image spot is still subwavelength for the first  "sphere" with $1.42<n_1<1.49$ and second sphere $n_2>1.5$ but $w/\lambda$ increases until $w=0.4\lambda$. The material with $n_1<1.42$ does not allow us to simulate the subwavelength spot even for $R/\lambda=(15-25)$. 
For $R/\lambda=30$ we need $n_1\ge 1.55$ and $n_2\ge 1.6$ so that to obtain $w/\lambda<0.5$.

\section{Subwavelength resolution of two dipoles}

\begin{figure}[htbp]
\centering

\includegraphics[width=0.48\textwidth,height=0.2\textwidth]{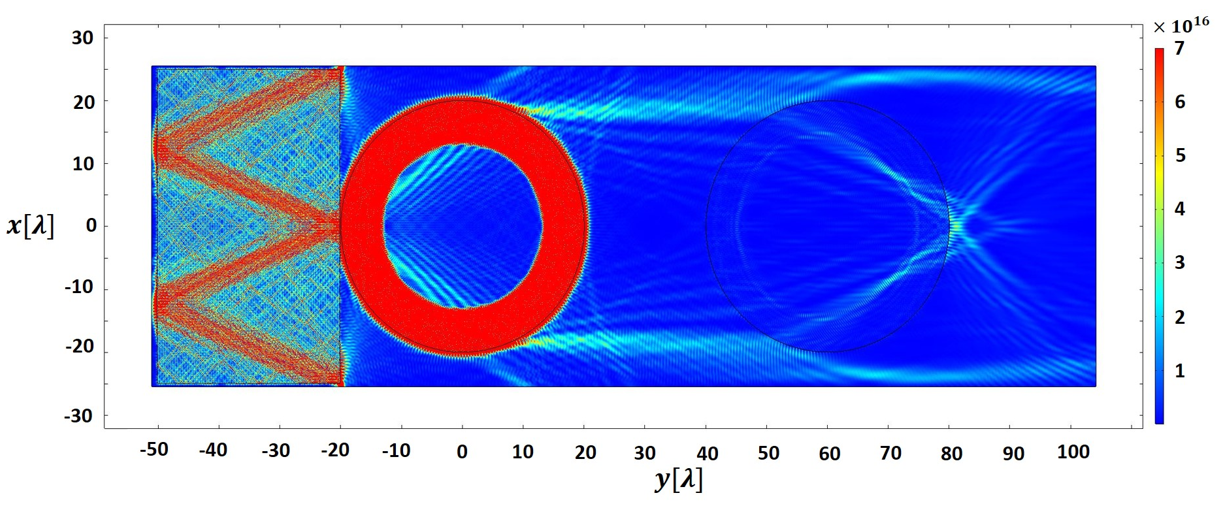}
\caption{Color map of the electric intensity in the imaging system repeating that from Fig.~\ref{Pic1} with added block of silicon 
distanced by $\lambda/10$ from the dipole source.}
\label{Pic11}
\end{figure}

Spatial resolution of two dipole sources is not always determined by {the effective focal spot ($w$) corresponding to single} point source. {It is proved in physical optics that the lateral resolution $\delta_{\rm min}$ of two mutually incoherent dipoles and the width of the single source imaging spot $w$ are equivalent}. However, this identity refers only to the case of two dipoles in free space, when one may express $w$ through the angular size of the Airy disk. {When two point sources, separated by a small gap $\delta$, are coupled to a microsphere or any other body, there is no equivalence of $w$ and $\delta_{\rm min}$.} Though these values are somehow related, we even cannot {be sure} that $w$ and $\delta_{\rm min}$ are values of the same order. 

Therefore, it is necessary to perform the simulations with two mutually incoherent dipoles. 
We have repeated the above simulations replacing one dipole source located at the point $y=0$,{ $x=-R-\lambda/10$} by 
two point dipoles stretched along $y$ and located at the points $y=\pm\delta/2$,{ $x=-R-\lambda/10$}. 
The gap $\delta$ is incomparably smaller than $R$, and these dipoles can be considered polarized radially to the imaging "sphere" center. The absence of mutual coherence of the dipoles is the same as the suppression of their interference. 
Since in our simulations, the oscillations are monochromatic, we could not mimic the mutual incoherence of two dipoles exactly. We emulate it by the phase shift $\pi/2$ between the dipole moments per unit length of our dipole lines. The fields of such dipoles do not interfere at least on the axis $y$, and in the region of our interest their interference is maximally suppressed. 

Our simulations have shown that $\delta\ge w$ in all cases when the imaging beam is non-divergent. For the most part of the explored ranges of two main parameters ($R=10-40\lambda$, $n=1.4-1.7$) the subwavelength resolution is not achieved, though the subwavelength imaging spot was obtained for an individual dipole. This is so, due to two reasons. First, the interference of two sources is suppressed not completely even in the paraxial region. Second, the GO approximation is not fully adequate for our "spheres" due to the impact of creeping waves, especially when $R<20\lambda$. Though the patterns of creeping waves are not visible in the simulated color maps, for $R<20\lambda$ their presence can be detected in the data files. 

However, in the presence of the substrate, the creeping waves are suppressed, and the GO model becomes much more adequate. It explains why the lower bound of the optimal interval $R/\lambda$ in the experimental studies is as low as $2.5\lambda$. The substrate broadens the range of $R/\lambda$ suitable for the formation of the imaging beam.

\begin{figure}[htbp]
\centering
\includegraphics[width=0.42\textwidth,height=0.29\textwidth]{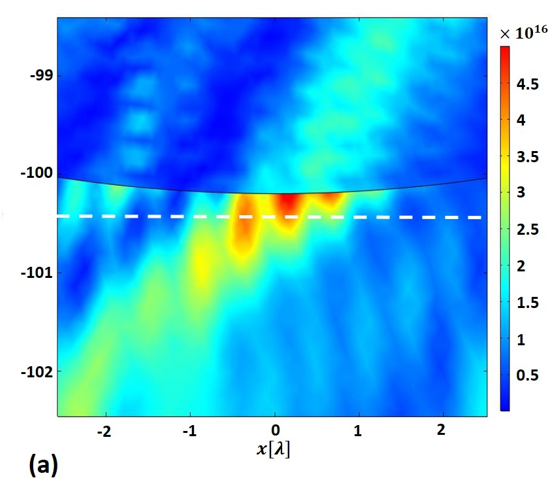}
\includegraphics[width=0.42\textwidth,height=0.29\textwidth]{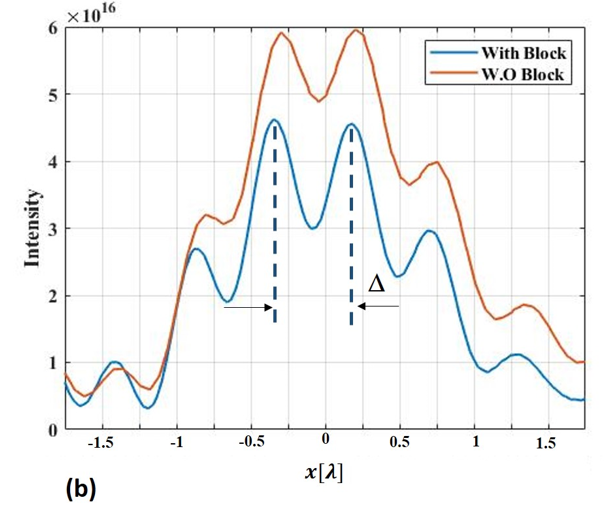}
\caption{Color map (a) and intensity plot in the focal plane (b) for the 
imaging system from Fig.~\ref{Pic11} excited by two dipoles with a gap {$\delta=0.27\lambda$} between them.
In the image this gap is magnified ($\Delta=1.93\delta$).  
}
\label{Pic4}
\end{figure}

We could not simulate the imaging structure with a macroscopic substrate. In the present work, we replace a realistic substrate by a substantial dielectric block of non-resonant sizes. The resonances of the block are excluded adding the optical absorption. {With such a block, we have managed to achieve a subwavelength resolution $\delta_{\rm min}\ll\lambda$ in the most part of simulated cases. }In presence of the block, the dipoles are sandwiched between its flat surface and the sphere. The minimal distance between the sphere and the block is adopted to be equal $\lambda/10$. 

First, let us see how the block changes the point dipole image size $w$. 
Fig.~\ref{Pic11} corresponds to the same imaging system as that depicted in Fig.~\ref{Pic1} with {one} difference -- a block of amorphous silicon added so that the dipole source is {sandwiched between} the block and the "sphere". 
In this case the intensity in the imaging beam mimics {much better} the intensity distribution calculated using the GO,
because the block completely eliminates the cavity modes. With the block we have achieved $w/\lambda=0.186$ instead of $w/\lambda=0.220$.

\begin{figure}[htbp]
\centering

\includegraphics[width=0.42\textwidth,height=0.29\textwidth]{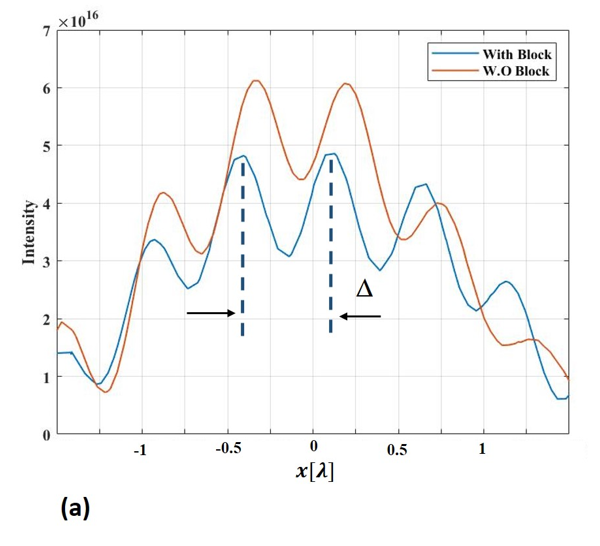}
\includegraphics[width=0.42\textwidth,height=0.29\textwidth]{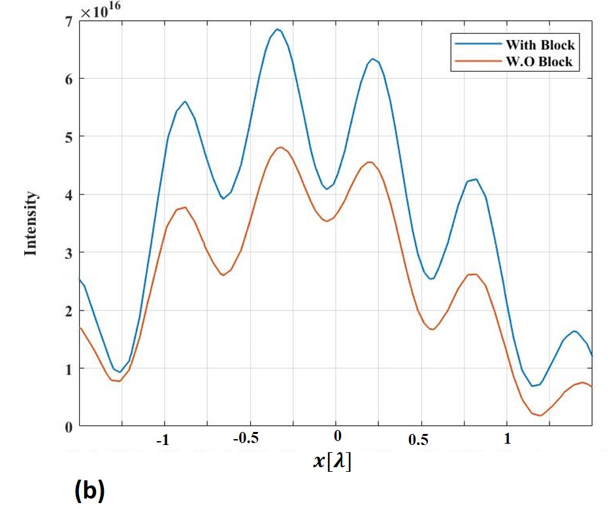}
\caption{ Intensity plots with and without the silicon block for the imaging system $R_1=R_2=30\lambda$, $n_1=n_2=1.6$ (a) and the same for 
the case $n_1=n_2=1.5$. In the last case, the resolution is not subwavelength because the refractive index of second sphere is not sufficient 
for a so large imaging "sphere". }
\label{PicNew3}
\end{figure}

Second, the presence of the block reduces $\delta_{\rm min}$, approaching it to $w$. 
In Fig.~\ref{Pic4}(a) we present the intensity color map in the area of the focal plane of the system $R_1=R_2=20\lambda$, {$n_1=1.525$ and $n_2=1.575$,} when the system is excited by two point dipoles located and polarized as it is explained above. The gap between them is as small as {$\delta=0.27\lambda$}.  The same dielectric block as in Fig.~\ref{Pic11} {results} in the clear separation of two intensity maxima in the focal plane. Note, the asymmetry of these maxima (different length along $y$) results from non-negligible interference of the fields produced by our dipoles beyond the axis $y$. However, in the focal plane, the intensity plot is almost symmetric. 
This plot is presented in Fig.~\ref{Pic4}(b) for both cases -- with and without the block. We see that in absence of the block, our dipoles are not {resolved according} to the Rayleigh criterion, but are {resolved in presence of it}. The gap $\delta$ is imaged in the focal plane by the distance $\Delta$ between two maxima. The image magnification $\Delta/\delta\approx 1.93$. Note that it is much smaller than the magnifications reported in the above-cited experiments with the imaging microspheres. We see two evident reasons for it. First, we simulate not a 3D microsphere but its 2D analogue. 
Second, we use a microlens located at a sub-mm distance from the sources and not a lens system of a microscope located much farther. The fact that the image magnification has been achieved, in principle, is more important. It shows that our second "sphere" operates as a lens, indeed.

The impact of the block is also positive for the regime of slightly diverging imaging beam. If we add a block to the imaging system $R_1=R_2=30\lambda$, $n_1=n_2=1.6$, excited with only one source, the imaging spot shrinks from $w=0.308\lambda$ to $w=0.275\lambda$. If we consider two sources with a lateral gap $\delta= 0.327\lambda$ and out of phase ($\pi/2$ shift), we can resolve them respecting the Rayleigh criterion only in presence of the block. Magnification in this case is $\Delta/\delta\approx 1.52$. In Fig.~\ref{PicNew3}(a) we present the intensity plots in the focal plane of the imaging system $R_1=R_2=30\lambda$, $n_1=n_2=1.6$ with and without the block. If we remove the block, the images of the similarly distanced point dipoles are not resolvable. However, for a so large "sphere" the subwavelength resolution even with the block holds only for high refractive indices $n_1=n_2\ge 1.6$. In Fig.~\ref{PicNew3}(b) we depict the similar intensity distributions for the case $n_1=n_2=1.5$. In this case two sources are distanced by {$\delta= 0.545\lambda$} and even a so substantial gap is not resolvable if we remove the block. Even though the block plays the positive role the truly subwavelength resolution is not achievable for a so small refractive index and large radius. 
These observations qualitatively fit the data from \cite{Hong} if we assume that in what concerns the formation of the imaging beams there is the correspondence between our 2D "spheres" of radius $R$ and 3D spheres of triply smaller radius.
In \cite{Hong} the spheres with $R=10-12\lambda$ offered the deeply subwavelength image ($\delta=\lambda/8$) when their refractive indices were as high as $n=1.7-1.8$, whereas the glass sphere ($n=1.46$) with so large radius did not offer the deeply subwavelength image. Meanwhile the glass spheres of smaller radius $R=5-7\lambda$ offered in these experiments the deeply subwavelength images.

\section{Discussion and conclusions}

In this article, we have assumed and partially validated (by exact numerical simulations) several scenarios of the non-resonant
and non-coherent far-field magnified subwavelength imaging enabled by a simple glass microsphere. We have shown that this functionality demands the object polarization normally to the sphere surface. The phenomenon is mainly 
based on the geometrical optics and the dipole radiation pattern, but it may involve also the beam Abbe diffraction if this diffraction transforms the beam into a nearly spherical wave. 
Besides of the partial theoretical validation of this scenario, we completely proved by exact numerical simulations that the imaging beam formed in accordance to the laws of GO 
keeps the subwavelength information on the object (located on the surface of a 2D microcylinder) at least up to the distances of the order of $100\lambda$. We have shown that such the object (consisting in our study of two non-coherent point dipoles) can be imaged by the second 2D microsphere with the subwavelength resolution and modest magnification.  

In this work we have found a range of the microsphere parameters for which the GO model is qualitatively adequate. We have determined the suitable intervals of these parameters for two scenarios of the far-field magnified imaging which supposedly correspond to the known experiments with the microspheres. We have shown: if the object is sandwiched between the substrate and the microsphere, the imaging functionality of the system of two microspheres improves. 

Some readers may by puzzled how the GO model may result in the superresolution, if this phenomenon is presumably related with the evanescent waves. However, evanescent waves are not necessary for label-free subwavelength imaging, it is a scientific myth which was demystified by a hyperlens. Inside the hyperlens an imaging beam propagates from a point-wise object towards the outer interface. The imaging beam keeps the fine information about the object location because it is thin compared to the wavelength in free space, however, this beam does not contain evanescent waves, it is formed by spatial harmonics which are all propagating in the hyperbolic medium. The imaging beam transmitted from the hyperlens to free space also does not contain evanescent waves, it is wider than $\lambda$, however, its spatial harmonics (propagating in free space) nicely 
keep the subwavelength information on the  location of the object. 

The same physics is responsible for the imaging by a simple glass microsphere. Our imaging beam also carries the fine information of the exact location of a point-wise source. Though our imaging beam is not subwavelength thin even in the zone of its waist, the information of the exact location of the object is in the direction of our beam. It offers the subwavelength accuracy for the dipole source location not  
due to the small thickness of our beam but due to the pronounced beam axis. This one exactly points out the location of the radially polarized dipole on the microsphere surface. This information 
is evidently finer compared to that offered by a subwavelength small thickness of the imaging beam propagating across the hyperlens. The axis of the hyperlens imaging beam is not as pronounced 
as the axis of our hollow imaging beam. This is one of the reasons why the imaging functionality of a simple glass microsphere is much better than that of the hemispherical hyperlens.
Another reason is the presence of plasmonic losses in the hyperlens, and the practical absence of losses in the optical glass. And the third reason is the narrow operation band of the 
hyperbolic metamaterial, whereas the glass keeps transparent in a very broad band of wavelengths. Though the hyperlens is made of a super-expensive nanostrucrtured metamaterial, 
the glass microsphere excited by a radially polarized object gains this contest.

To conclude: we think that the transition from the 2D "sphere" to a 3D sphere promises a finer subwavelength resolution than that we have obtained in the simulations of this paper. And we believe that this resolution and magnification predicted by our model will be much higher when the macroscopic objective lens is used instead of the microlens and the distance from the imaging microsphere to the objective is realistic -- corresponding to the cited experiments. We believe that our work is important as a conceptual proof of the non-resonant far-field subwavelength imaging functionality of the dipole radiation transmitted through the microsphere in accordance to the simplest geometric-optical model.

%\section{Acknowledgement}
%C. Simovski acknowledges the support of Russian Science Foundation (Project №21-79-30038).
%Funding by the Russian Foundation for the Basic Research (Project №21-79-30038) is acknowledged by V.K.
%\section{Disclosures}
%The authors declare no conflicts of interest.


\bibliographystyle{model1-num-names}

\bibliography{paper}

\end{document}